\documentclass[aip,cha,reprint]{revtex4-1}
\usepackage{amscd}
\usepackage{amssymb}
\usepackage{amsfonts}
\usepackage{amsmath}
\usepackage{graphicx}
\usepackage{dcolumn}
\usepackage{bm}
\usepackage{multirow}
\usepackage[dvipdfm]{hyperref}
\usepackage[percent]{overpic}%

\begin{document}

\title{Exploration of period-doubling cascade route to chaos with complex network based time series construction}

\author{Ruoxi Xiang}
\email{rxxiang@gmail.com}
\affiliation{Department of Electronic and Information Engineering,
Hong Kong Polytechnic University, Hong Kong, People's Republic of
China}

\author{Michael Small}
\affiliation{School of Mathematics and Statistics, University of Western Australia, Crawley, WA 6009, Australia}

\date{\today}

\begin{abstract}
In this work,  the topologies of networks constructed from time series from an underlying system undergo a period doubling cascade have been explored by means of the prevalence of different motifs using an efficient computational motif detection algorithm. By doing this we adopt a refinement based on the $k$ nearest neighbor recurrence-based network has been proposed. We demonstrate that the refinement of network construction together with the study of prevalence of different motifs allows a full explosion of the evolving period doubling cascade route to chaos in both discrete and continuous dynamical systems. Further, this links the phase space time series topologies to the corresponding network topologies, and thus helps to understand the empirical ``superfamily'' phenomenon, as shown by Xu.
\end{abstract}


\maketitle

\section{Introduction}
 Networks are considered to be abstract models which simply focus on elements within complex systems and the interactions among them. Yet, structurally they could be very complex. To understand the topological structure of a complex network is important because it usually holds the key to the secret of how a complex system functions. Specifically, for network representatives from time series~\cite{Xu:Superfamily:2008,Zhang:Complex:PhysRevLett.96.238701,Reik:Recurrence:2010,Lacasa:From:2008,Yang:Complex:2008,Gao:Complex:2009,Donner:Ambiguities:2010,shimada:analysis:2008,xiang:multiscale:2012,sun:characterizing:2014}, the information of time series is mapped into network structures and the key question becomes what kind of information of the time series is contained in the corresponding networks and how it manifests within the networks. The answer differs depending on the what kind of construction methods we use to transform time series into networks and hence we need to develop proper mapping methods so that the network could best preserve information of the time series. After a proper transformation, the investigation to the characteristics of time series can be performed thanks to the rapid development of complex network theory, and, in return, networks descriptions provide new perspectives to time series.
 
To constructing networks from time series data, a variety of approaches have been proposed and these methods can be roughly been divided into three families: proximity networks, transition networks and visibility graphs~\cite{Donner:Recurrence:2011}. As they named, proximity networks are constructed by making use of the mutual proximity of segments of data~\cite{Zhang:Complex:PhysRevLett.96.238701,Yang:Complex:2008,Gao:Complex:2009,Reik:Recurrence:2010,Xu:Superfamily:2008,xiang:multiscale:2012}, while transition networks are built based on the transition connections between states~\cite{nicolis:dynamical:2005,shirazi:mapping:2009,campanharo:duality:2011}. Unlike the above two classes of networks, visibility graphs are distinct for they are constructed by making use of the convexity relationships of the data~\cite{Lacasa:From:2008}.

Recognized as the the biggest family, proximity network methods ---  in particular the two types of recurrence based networks (that are $k$ nearest neighbor networks and $\varepsilon$-recurrence networks which have provided new perspectives to the attractor properties in the phase space~\cite{Donner:Recurrence:2011,Donner:geometry:2011,donner:recurrencenjp:2010,Donges:Functional:2012,xiang:multiscale:2012}) arouse intensive research interest. Specifically, in the $k$ nearest neighbor phase space networks~\cite{Xu:Superfamily:2008}, data of different dynamics from both discrete and continuous systems can be distinguished into different groups and exhibit the so-called superfamily phenomenon when studying the occurrence subgraphs of order 4 . Detailed characterizations of such networks have been studied~\cite{xiang:multiscale:2012}. The network characterizations that have been discussed cover not only the global summary statistics like the average path length, clustering coefficient and degree distribution, but also measurements of a specific node such as its node degree, local clustering coefficient and betweenness centrality. From the complex networks' point of view, these network characterizations explain network structure at different scales, no matter if it refers to global  statistics or the local node properties, and hence provide profound information about the original time series. In particular, more detailed micro- and meso- scale structural features around a specific node are contained in the local node properties which can provide more detailed information of the phase space attractor that is otherwise buried in the average geometry of the attractor. Also,  the effect of noise to the so-called superfamily phenomenon suggested by Xu {\em et al.}~\cite{Xu:Superfamily:2008} has been examined by means of calculating the ranking of motif frequencies of the networks that are constructed from real-world data.

Regarding the network motifs, which are defined as basic building elements in terms of small connected subgraphs in a network which occur in a particularly higher frequencies compared which those which would be expected in a randomized network\cite{Milo:Network:2002}, they have recently attract much research interest. When building complex network models using specific structural design principles, some types of subgraphs would occur with a higher frequencies than in random networks and hence network motifs give insights beyond the trivial scale of just individual nodes and links. Therefore the analysis of motifs can lead to meaningful results of the mechanism of how a complex network taking shape and how the original system functions. There have been many interesting results obtained by using network motif analysis. Examples can be found in a variety kind of real world networks, including electronic circuits and power distribution networks, ecological networks, software engineering diagrams, World Wide Web, biological networks and social networks. Also, there are more related works done by analysis of motifs in networks that are constructed from time series~\cite{Xu:Superfamily:2008,Dong:network:2013,Gao:motif:2010,Li:Detection:2011}.

 The motif ranking is a global statistics. Rather than the ranks of motifs, we believe that the local motif information, which refers to the numbers of different types of subgraphs which a given node $v$ belongs to in the network, can provide more detailed local phase space properties. So far, the order $s$ of motifs in networks constructed from time series that has been studied is chosen to be $s=4$, for the reasons that with $s=4$, it provides enough essential network information needed and that with an acceptable computational cost. Nevertheless, changing the size of motifs $s$, local network topologies at a changeable scale is obtained. Studying the occurrence frequencies of the local motif patterns can help to understand how they reflect the local temporal structure of a phase space of the time series and thus explain the emergence of the ``super family'' phenomenon. Hence, this work focus on the concepts of network motifs, especially the local motifs, aiming to link the motifs with the topologies of time series.

The reminder of this work proceeds as follows: we introduce a method for motif detection, followed by the modified methodologies for the construction of $k$ nearest neighbor phase space network which are more suitable for motif calculation. Next, we provide analysis obtained by applying motif detection methods and network cotransduction methods to numerical simulations.

\section{\label{sec2:}Methodology}
This section presents the methodologies including tools that we use to detect network motifs and the way of how we build the complex networks from time series.
\subsection{Tools for motif detection}
There are many tools that are designed to find motifs in complex networks, publically available implementers include Kavosh, Mfinder, MAVisto, Pajek and FANMOD~\cite{Kashani:kavosh:2009,Kashtan:efficient:2004,Schreiber:mavisto:2005,Batagelj:pajek:2004,Wernicke:fanmod:2006}. Detecting network motifs usually can be divided into three computationally challenging subtasks:
\begin{enumerate}
\item \textbf{Enumeration} Look for all subgraphs of a given size in the network.
\item \textbf{Classification} Classify and group subgraphs which are isomorphic accordingly.
\item \textbf{Random graph generation} Determine the significance of the subgraphs by comparing with random network.
\end{enumerate}

\begin{figure}[!ht]
 \centering
\includegraphics[trim=0 0 0 0, clip=true, scale=0.3]{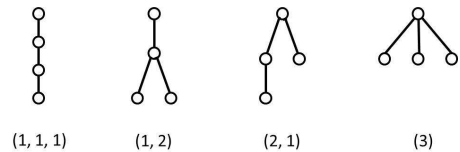}
\parbox{8cm}{\caption{\label{fig:tree}
Four types of trees regarding all the four possible compositions of $s-1=3$.}}
\end{figure}

Regarding the first subtask --- to identify all the subgraphs in the network --- we choose the Kavosh algorithm~\cite{Kashani:kavosh:2009} in this work for its efficiency in subgraph enumeration which requests less CPU time, less memory usage, and capability to deal with motif of a larger size. The enumeration of subgraph of size $s$ of Kavosh algorithm starts with building trees of depth $m$ and size of $s$ rooted at a particular node of the network. To descend the tree, a child node is eligible to be chosen only if it does not in any upper level of the tree (i.e. it has not yet been visited). The selected nodes are marked as visited and again the tree is descended. Once the tree has reached to the lowest possible level, it is ascended and the nodes located in previous lower paths will be treated as unvisited again. The number of children $s_i$ to be chosen in level $i$ is determined by the all the possible composition operation of $s-1$ with $s_2+s_3+\ldots+s_m=s-1$. Take $s=4$ as an example, all the possible compositions include (1,1,1), (1,2), (2,1), (3). Hence we get all the possible types of trees for order 4 as show in Fig~\ref{fig:tree}.

Regarding the second subtask that is to identify graph isomorphism, a power algorithm called NAUTY~\cite{mckay:practical:1981} has been adopted in most of the motif determination tools. After all the possible trees rooted at a specific node with $k-1$ descendants are extracted, all the subgraphs which include that node are obtained. These subgraph will be given a unique identifier called \textit{canonical labeling} when using the so called NAUTY algorithm. The labeling is from a string by flattening the adjacent matrix. The \textit{canonical labeling} is defined as the smallest (or largest) string among those from flattening all matrices generated by permutating rows and lines of the adjacency matrix, as a graph remains the same if the ordering of nodes are switched. Subgraphs with the same canonical label are in the an isomorphism class and hence belong to a some particular motif type. We can obtain the local motif information of the root node, to be more precise, the number of different motif types corresponding to the root node which provides proud local information of that node. Besides, if one aims to know the global motif ranking information of the complex network, the present root node is removed from the network and we select another node in the network as a new root node and repeat the same procedure until all the subgraphs in the network are found.

We should note here no matter what kind of motif determination tools are used, the second subtask is usually very time consuming when the size of motif goes large, limiting the implement of motif determination tools for large size of motif detection.

\begin{figure}[!ht]
 \centering
\includegraphics[trim=0 0 0 0, clip=true, scale=0.5]{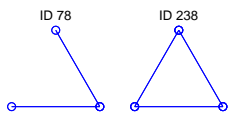}
\parbox{8cm}{\caption{\label{fig:Motiflabel3}
Motif dictionary of all the undirected subgraphs of order 3 with their canonical labels.
}}
\end{figure}

\begin{figure}[!ht]
 \centering
\includegraphics[trim=0 0 0 0, clip=true, scale=0.5]{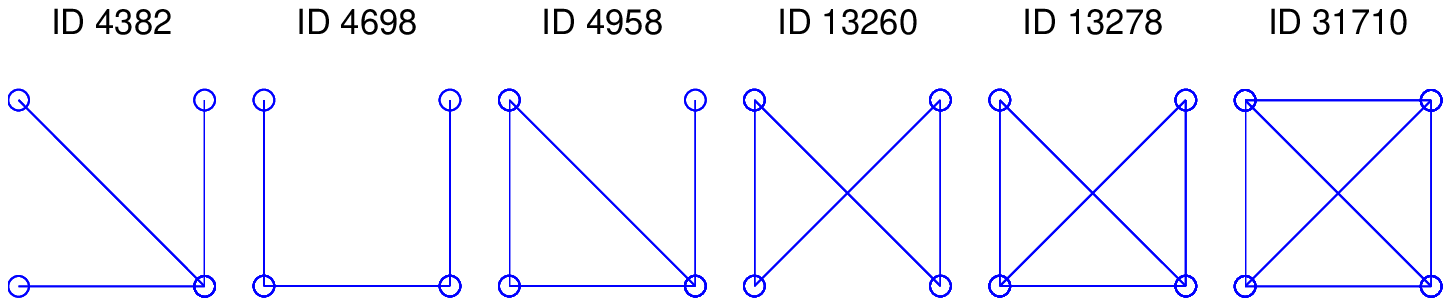}
\parbox{8cm}{\caption{\label{fig:Motiflabel4}
Motif dictionary of all the undirected subgraphs of order 4 with their canonical labels.
}}
\end{figure}

\begin{figure}[!ht]
 \centering
\includegraphics[trim=0 0 0 0, clip=true, scale=0.5]{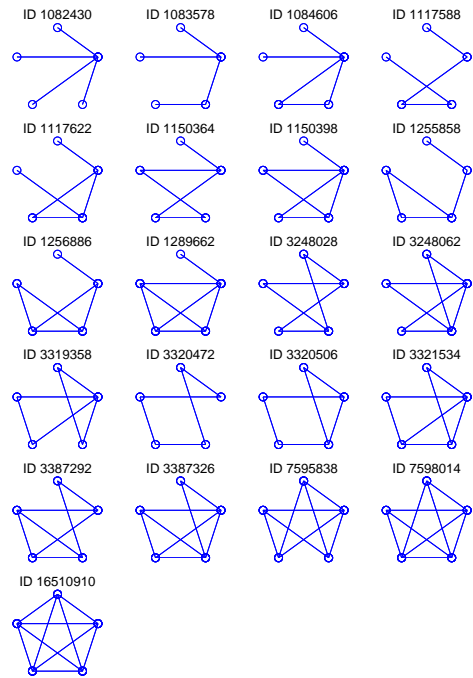}\\
\parbox{8cm}{\caption{\label{fig:Motiflabel5}
Motif dictionary of all the undirected subgraphs of order 5 with their canonical labels.
}}
\end{figure}

 For reference, the shape of all the undirected subgraphs of order 3, 4 and 5 with their canonical labels are shown in Figs.~\ref{fig:Motiflabel3}-\ref{fig:Motiflabel5}. Ref.~\onlinecite{Xu:Superfamily:2008} considers only motifs up to order 4 due to the heavy computational cost for calculating higher order of motifs. By using Kavosh algorithm, it is possible to look into motifs of a higher order in a network --- at least for some particular nodes of the network that we are interested in.


\subsection{\label{chap04:networkconstruction}Network construction}
The implementation of $k$-nearest neighbor phase space network construction for follows the method used in Refs.~[\onlinecite{Xu:Superfamily:2008}], [\onlinecite{Donner:Recurrence:2011}] and [\onlinecite{xiang:multiscale:2012}]. In more details, the time series is embed in an appropriate phase space and each point in the phase space is then taken as a node of the network. By comparing pairwise phase space distance among nodes, we can obtain an index matrix $I$ in which each element $I_{ij}$ is a node index representing the $j$th nearest neighbor of node $i$ is the node with index $I_{ij}$. In this work, we further refine the original algorithm of Xu by introducing a tiebreaker for those eligible neighbors in the index matrix which are the same distance apart from a given node when building the index matrix $I$. Such a situation could happen when the points in phase space are distributed uniformly,  e.g., reconstructed from a periodic data set, leading to some ambiguity in building networks. To avoid the ambiguity, for a given node, we start from its successors by assuming that point which ``follows'' the given node and at the same time temporally closer should have a higher priority among those neighbors. For example, let $x_i$, $x_j$, $x_m$, $x_n$, $x_p$, $x_q$ be six potential neighbors of node $x_o$ (which means they are not on the same orbit as $x_o$ with temporal relationship of $t_i<t_j<t_m<t_n<t_o<t_p<t_q$ and with the pairwise distance relationship of $d_{oi}<d_{om}=d_{on}=d_{op}=d_{oq}<d_{oj}$ to node $x_o$. The priority in ranking should be $i$, $p$, $q$, $m$, $n$, $j$. Furthermore, we adopt the periodic boundary condition as a compensation  for those who locate temporally behind.

\begin{figure}[!ht]
 \centering
\includegraphics[trim=0 0 0 0, clip=true, scale=0.4]{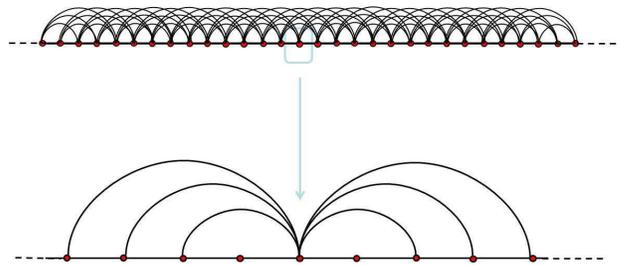}
\parbox{8cm}{\caption{\label{fig:networkperiodic}
Typical network structures which are commonly found in networks from the simplest data sets of constant values ($k=4$). Each node is connected to its $2k$ nearest neighbors, namely to the $k$ nearest neighbors clockwise and counter-clockwise respectively.}}
\end{figure}

Note that as proposed by Xu {\em et al.}, when building the index $I_{ij}$, the points lie on the same trajectories are excluded as eligible neighbors and will not be shown in the index matrix.  On the one hand this ensures that the neighbors are spatial neighbors in the phase space rather than the temporal neighbors, on the other hand this avoids multiple links between two nodes.  Compared with the original algorithm, the improvement seems apparently minor but it ensures that networks constructed from periodic data sets could be regular in topology and exhibit uniformly distributed network properties, as expected (see Fig.~\ref{fig:networkperiodic} as an example), and thus ensures the occurrence of motifs can represent typical dynamics of the data.

\section{Applications to data from dynamical models}
In this section, we apply both motif detection method and network construction methods to data from the logistic map and the R\"{o}ssler system which are taken as benchmark examples of map data and flow data. We use the Kovash algorithm to look for subgraphs of a given size in a network and label them by classifying each of them into different motif types by applying the NAUTY algorithm. Note that as the network construction methods already guarantees that the $k$-nearest neighbor phase space networks are with a same edge density as the size of the network is fixed, we do not have to perform random network generations to compare the subgraphs frequencies with random networks since we focus only on the difference among networks construction from time series of different dynamical properties. Before moving to the two toy models, we first apply the methods to the simplest regular infinite networks

\subsection{Regular infinite networks}

As a case study, we apply Kavosh algorithm to regular infinite networks that can be obtained from data sets of constant values. Every node in a regular network has the same property. Hence we can obtain the absolute number of motifs of different types for every node and that can be a standard when make comparison among networks from different dynamics. The orders of subgraphs $s$ we investigate range from 3 to 5 as listed in Tables~\ref{table:regular3}-\ref{table:regular5}. In each table, we set $k$ as 5, 4, 3, 2 and 1 in each column respectively and the total number of subgraphs are listed in the last row. It is easy to find that the total number of subgraphs of a node can be expressed as $s\cdot k^{s-1}$. The frequencies of subgraphs (no matter the frequencies of individual type of subgraphs or the total frequencies of subgraphs) of the whole networks are proportional to that of a single node and the number of nodes of the network.

\begin{table*}[!h]
\caption{\label{table:regular3}
Subgraph frequencies of order $s=3$ for a node in a regular network with $2k$ neighbors.}
\begin{center}
\begin{tabular}[c]{rrrrrr}\hline\hline
Motif ID            & $k=5$ & ~~4   & ~~3    & ~~2    & ~~~1    \\\hline
78                  & 45    & 30    & 18     & 9      & 3       \\
238                 & 30    & 18    & 9      & 3      & 0       \\\hline
Total  frequencies  & 75    & 48    & 27     & 12     & 3       \\\hline\hline
\end{tabular}
\end{center}
\end{table*}

\begin{table*}[!h]
\caption{\label{table:regular4}
Subgraph frequencies of order $s=4$ for a node in a regular network with $2k$ neighbors.}
\begin{center}
\begin{tabular}[c]{rrrrrrr}\hline\hline
Motif ID  & ~~Type& ~$k=5$ & ~~~4     & ~~~~3      & ~~~~2      & ~~~~~1       \\\hline
4698    &A      & 220   & 120     & 56       & 20       & 4       \\
4958    &B      & 160   & 80      & 32       & 8        & 0       \\
13278   &C      & 80    & 40      & 16       & 4        & 0       \\
4382    &D      & 0     & 0       & 0        & 0        & 0       \\
13260   &E      & 0     & 0       & 0        & 0        & 0       \\
31710   &F      & 40    & 16      & 4        & 0        & 0       \\\hline
Total  frequencies &   & 500   & 256     & 108      & 32       & 4 \\\hline\hline
\end{tabular}
\end{center}
\end{table*}

\begin{table*}[!h]
\caption{\label{table:regular5}
Subgraph frequencies of order $s=3$ for a node in a regular network with $2k$ neighbors.}
\begin{center}
\begin{tabular}[c]{rrrrrrr}\hline\hline
Motif ID     & ~$k=5$ & ~~~4     & ~~~~3      & ~~~~2      & ~~~~~1       \\\hline
1117588         & 950    & 425     & 155       & 40       & 5       \\
1117622         & 175    & 75      & 25        & 5        & 0       \\
1255858         & 850    & 350     & 110       & 20       & 0       \\
1256886         & 500    & 200     & 60        & 10       & 0       \\
1289662         & 150    & 50      & 10        & 0        & 0       \\
3319358         & 75     & 25      & 5         & 0        & 0       \\
3321534         & 175    & 75      & 25        & 5        & 0       \\
3387326         & 150    & 50      & 10        & 0        & 0       \\
7598014         & 75     & 25      & 5         & 0        & 0       \\
16510910        & 25     & 5       & 0         & 0        & 0       \\\hline
Total  frequencies      & 3125   & 1280    & 405       & 80       & 5       \\\hline\hline
\end{tabular}
\end{center}
\end{table*}

Similar results can be obtained in circulant with  the jump sequence (1, 2, 3, ..., $k$) as the network size are large enough.

\subsection{\label{chap04:logistic}Logistic map}
Logistic map is one of the simplest maps, exhibiting complicated dynamics and forming the foundation of a universality class. It is expressed as
\begin{equation}
  x_{i+1} = \lambda x_i(1 - x_i)
\end{equation}
in which $\lambda$ is the control parameter, $0\leq\lambda\leq 4$, $x\in[0,1]$.

\begin{figure}[!ht]
\centering
\includegraphics[trim=0 0 0 0, clip=true, width=0.4\textwidth]{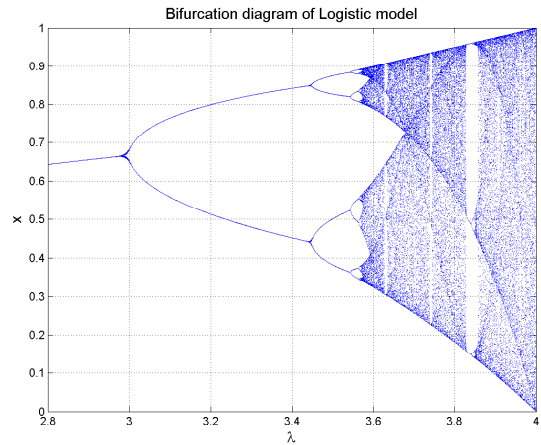}
\parbox{8cm}{\caption{\label{fig:bifurcation}
Bifurcation diagram, $x$ versus the control parameter $\lambda$.
}}
\end{figure}

When $\lambda$ is varied between 0 and 4, different types of dynamics are encountered in a complicated and intermingled way after the system getting rid of a transient period and the bifurcation diagram is shown in Fig.~\ref{fig:bifurcation}.

\begin{itemize}
\item With $\lambda$ between 0 and 3, $x$ will approach a fixed value  $\displaystyle\frac{\lambda-1}{\lambda}$.
\item With $\lambda$ between 3 and $\lambda_\infty$ ($\lambda_\infty\simeq3.5699456$), $x$ will approach oscillations of $2^i$ period limits cycles with $i=1, 2, 3...$ as $\lambda$ increases accordingly. Such behaviors are also called \textit{period-doubling bifurcation}.
\item At $\lambda_\infty$ the period-doubling cascade columnates with the onset of chaos. With $\lambda$ between 0 and 4, most values of $\lambda$ led to chaotic behavior but there are also period oscillation for some values of $\lambda$. At $\lambda=4$, there is one chaotic band for $x$ ranging from 0 to 1. As $\lambda$ decreases starting from 4, we can find the number of chaotic bands doubling at some values of $\lambda$. Conversely, as $\lambda$ increases, the number of chaotic bands halving and such behaviors are called \textit{period-halving cascade of chaotic bands}.
\end{itemize}

\begin{table*}[!h]
  \caption{\label{table:criticalpoints}
The values of $\lambda_{2^i}$ for the onset of $2^i$ period limits cycles and $\lambda_{(2^i)}$ for the onset of period-halving of $2^{i-1}$ chaotic bands}
\begin{center}
\begin{tabular}[c]{lrrrr}\hline\hline
$i$     &cycles($2^i$)/bands($2^{i-1}$)    &$\lambda$  &Description \\\hline
3       &8                      &3.544090359 &4 $\rightarrow$ \quad8\\
4       &16                     &3.564407266 &8 $\rightarrow$  $\;\:$16\\
5       &32                     &3.56875942  &16 $\rightarrow$  $\;\:$32\\
6       &64                     &3.56969161  &32 $\rightarrow$  $\;\:$64\\
7       &128                    &3.569891259 &64 $\rightarrow$ 128\\
8       &256                    &3.569934019 &128 $\rightarrow$ 256\\

$\infty$&\mbox{Accumulation point}&~~~~~3.569945672 &\mbox{The onset the chaotic region}\\\hline
6       &32                     &3.56999339  &64 $\rightarrow$ 32\\
5       &16                     &3.570168472 &32 $\rightarrow$ 16\\
4       &8                      &3.57098594  &16 $\rightarrow$ $\;\:$8\\
3       &4                      &3.574804939 &8 $\rightarrow$ $\;\:$4\\
2       &2                      &3.592572184 &4 $\rightarrow$ $\;\:$2\\
1       &1                      &3.678573501 &2 $\rightarrow$ $\;\:$1\\\hline
\end{tabular}
\end{center}
\end{table*}

\begin{figure}[!ht]
\centering
{\begin{overpic}[trim=0 0 0 0, clip=true, width=0.38\textwidth]{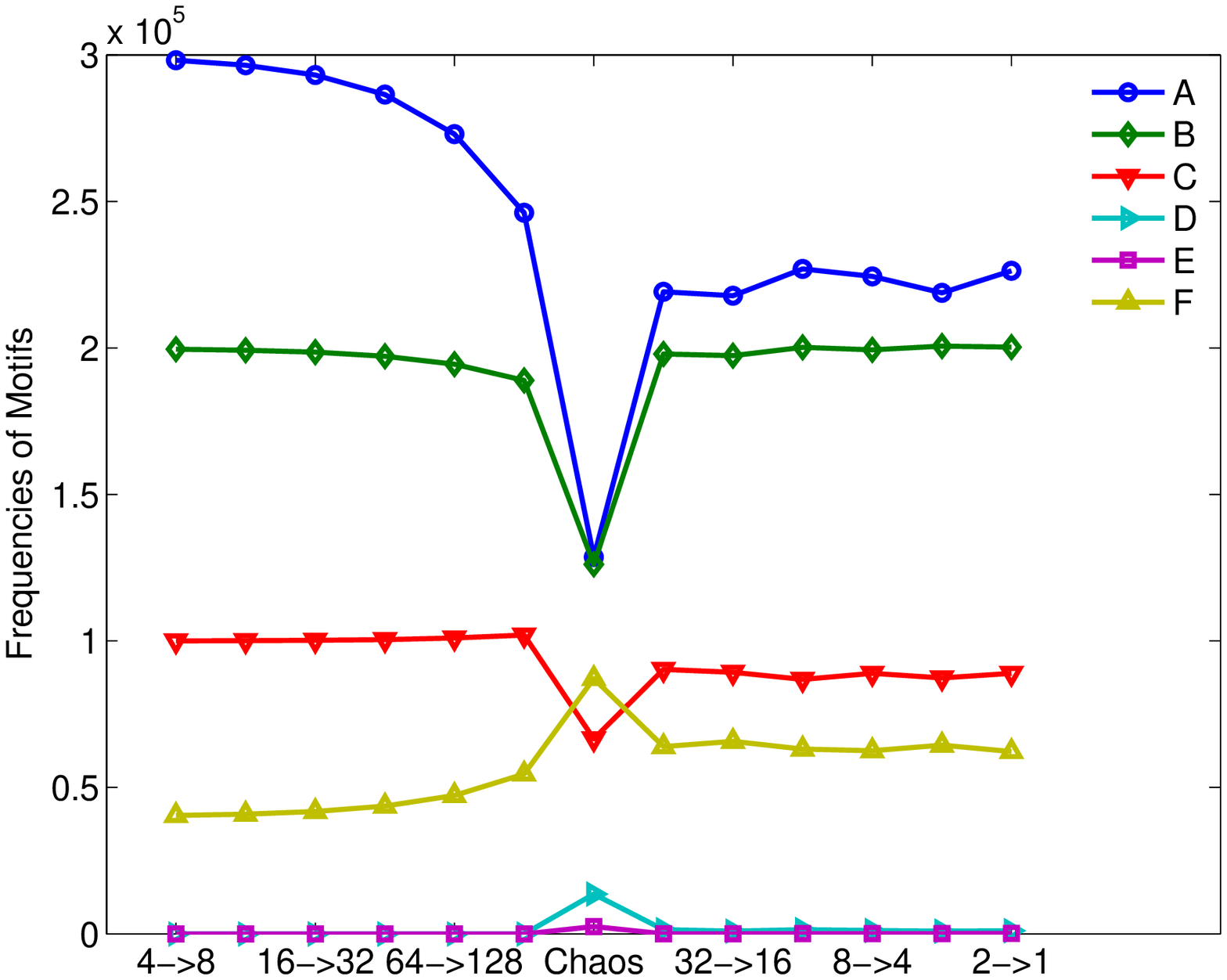}
\put(90,10){(a)}
\end{overpic}\\
\begin{overpic}[trim=0 0 0 0, clip=true, width=0.38\textwidth]{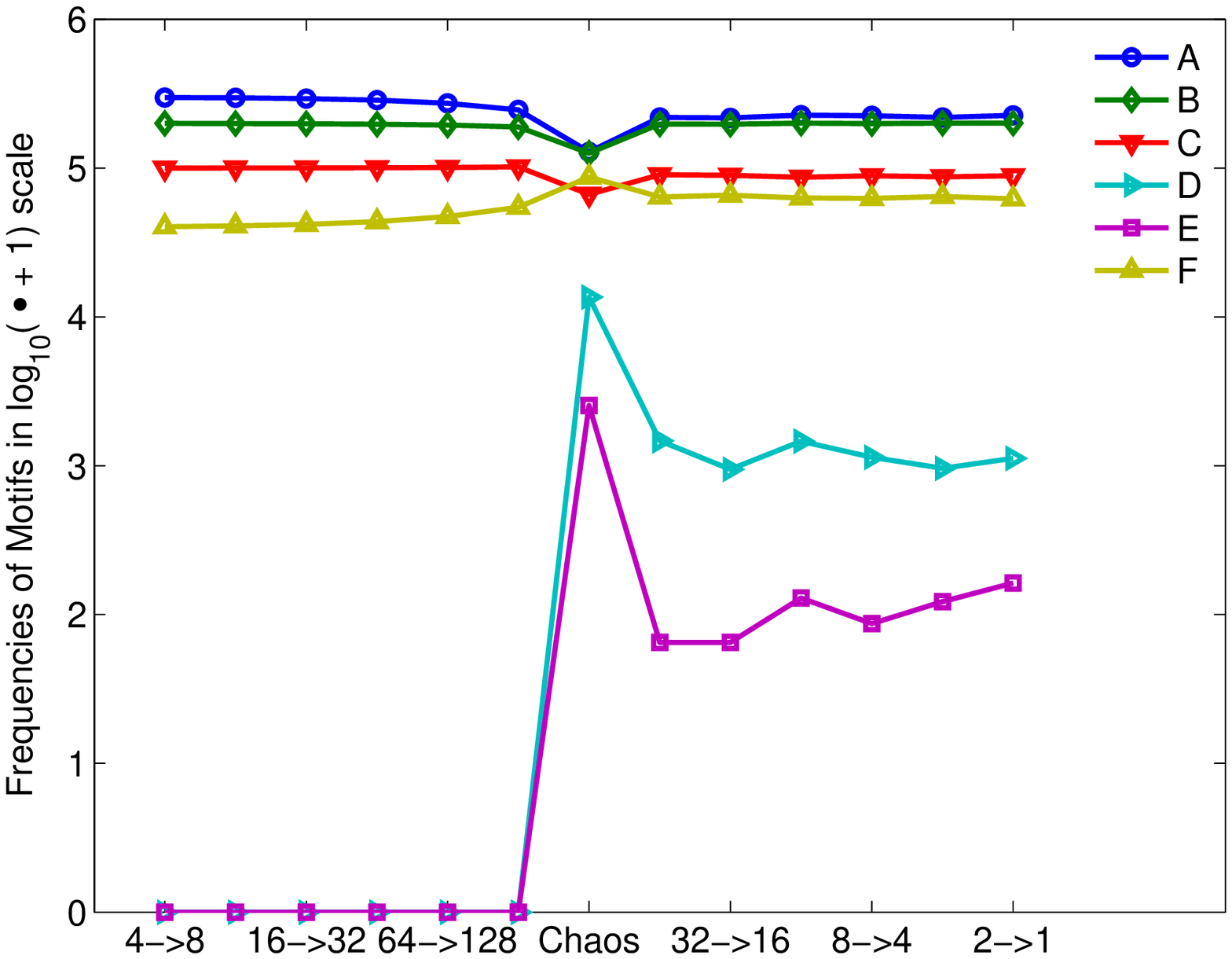}
\put(90,10){(b)}
\end{overpic}
}\\
\parbox{8cm}{\caption{\label{fig:Motiflogbif}
(a) Motif frequencies for both period-doubling route to chaos and period-halving of chaotic bands. (b) The same plot in logarithm scale.
 }}
\end{figure}

 We make use of data sets with different $\lambda$ to obtain phase space network of size $n=10000$. We embed each of the data sets into a phase space with the time delay $\tau=1$ and the embedding dimension $d_e=5$ and consider each point in the phase space as a node of the network. (Note that the size of the networks $n=N-\tau\cdot (d_e-1)$ are different from the data size $N$, due to the embedding procedure. This does not affect the main results since $\tau\cdot (d_e-1)$ are usually far less than $N$.) By making use of the mutual proximity of points in the phase space, we assign each point with $k=4$ nearest neighbors to get the phase space network following the modified methods that has been introduced in the previous section. We would like to see how the phase space network can capture the system's changing behaviors so we start with the critical points for the system's behaviors transitions of both period doubling and chaotic bands period-halving as shown in Table~\ref{table:criticalpoints}. Figure~\ref{fig:Motiflogbif} shows the motif frequencies in both linear and logarithm scale for each data set. The $x$ axis of the figure indicates that the increase in the values of $\lambda$ leads to corresponding changes in the system's dynamics accordingly. Hence, the first half of Figs.~\ref{fig:Motiflogbif} from $4\rightarrow8$ to chaos shows the results for period-doubling bifurcation, while the second half of the figure, that is from chaos to $2\rightarrow1$, shows the result for period-halving of chaotic bands.

 To begin with, it is clear that the motif frequencies of $A$, $B$, $C$, $D$, $E$ and $F$ for the onset of 4 to 8 period doubling are proportional to that of a single node in the regular network (see the $k=4$ case in Table.~\ref{table:regular4}, i.e., $120$, $80$, $40$, $0$, $0$, $16$). As the number of period increasing, we observe the decreasing trend in motifs $A$ and $B$, the slightly increasing trend in $C$ and $F$ as well as none of type $D$ or $E$. Thus can be explained by the topologies of the corresponding network in which the network is separated into an increasing number of components of almost equal size because of period doubling and, at the same time, each components contains less nodes connected together, hence the motif frequencies differs more from regular networks of infinite length as period increases. We can infer that if we increase the length of the data, the proportion of the frequencies of each kind of motifs will tend to that from regular networks of infinite length for $k=4$ case, as shown in Table.~\ref{table:regular4}. At the onset of the chaos region, we even observe the motif ranking switching between $C$ and $F$ as well as the emergence of motif of type $D$ and $E$. The motif frequencies for critical points of period-halving of chaotic bands are almost the same. All the cases of period-halving of chaotic bands belong to a same superfamily, i.e. $ABCFDE$.  Regarding the absolute values, we found decreases in motifs $A$ and $C$, as well as increases in motifs $B$, $D$, $E$, and $F$, compared with that in the periodic doubling cascade respectively.   As we could expect, the decrease in frequencies of motifs $D$ and $E$ implies that the attractors become more heterogeneous. Conversely, for motif $F$, which would be expected to be more prevalent when more points are evenly distributed in low dimensional attractors, occurs more often for points in chaotic attractors compared with periodic attractors. The reason for this change could be that, in period-halving of chaotic bands cases, the networks could be divided into components in uneven sizes with heterogenous connections. There could exist small but densely connected components or clusters in the networks, leading to the increase in motif $F$ comparing to regular networks, as well as the decrease of the most dominated motif of type $A$. Note that when building $k$-nearest neighbor phase space network for flow data, if an incommensurate sampling frequency is chosen, the network could also be divided into unconnected components, leading to unexpect values in network measurements for some realizations.

\begin{figure}[!ht]
\centering
{\begin{overpic}[trim=0 0 0 0, clip=true, width=0.38\textwidth]{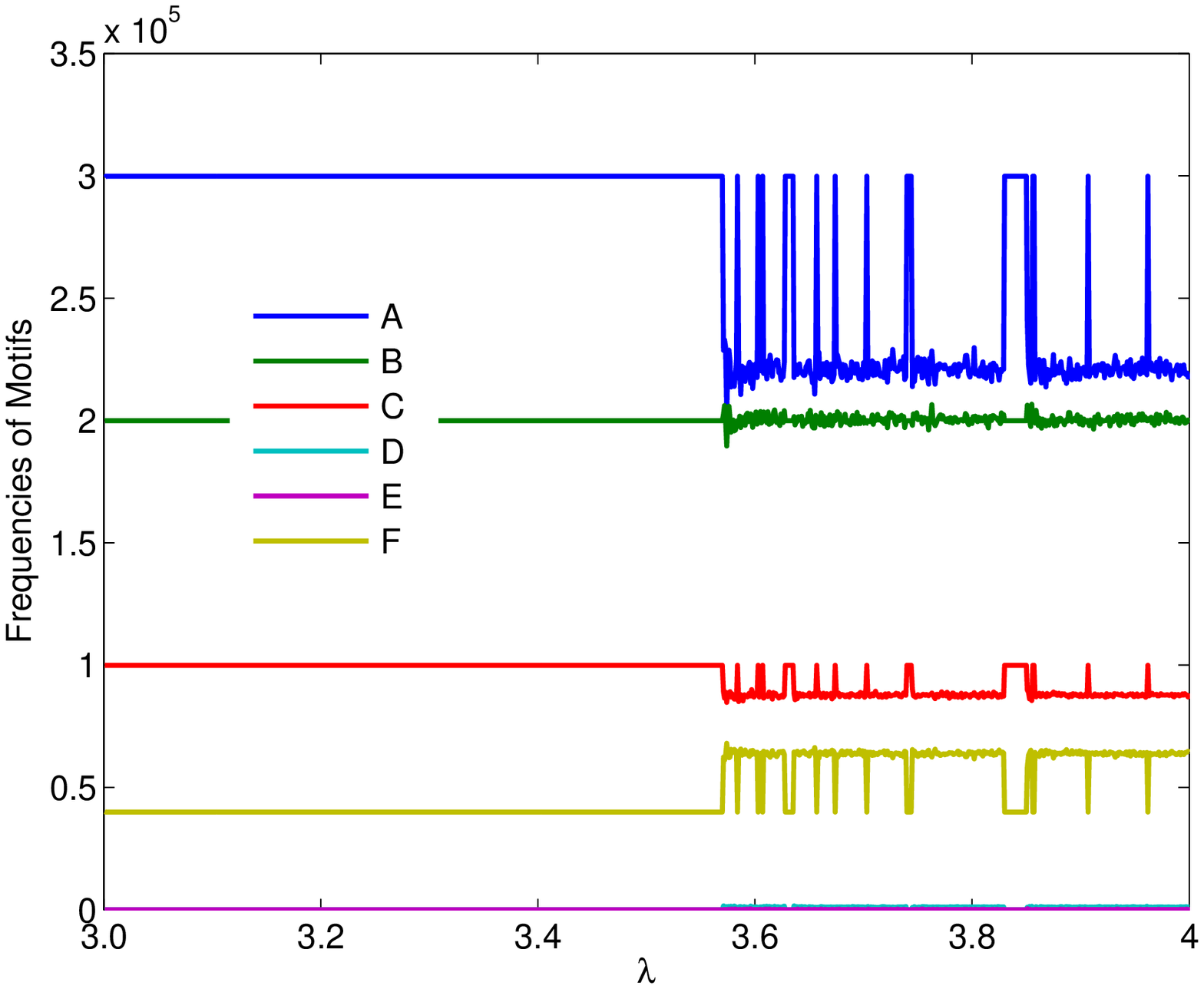}
\put(12,10){(a)}
\end{overpic}\\
\begin{overpic}[trim=0 0 0 0, clip=true, width=0.38\textwidth]{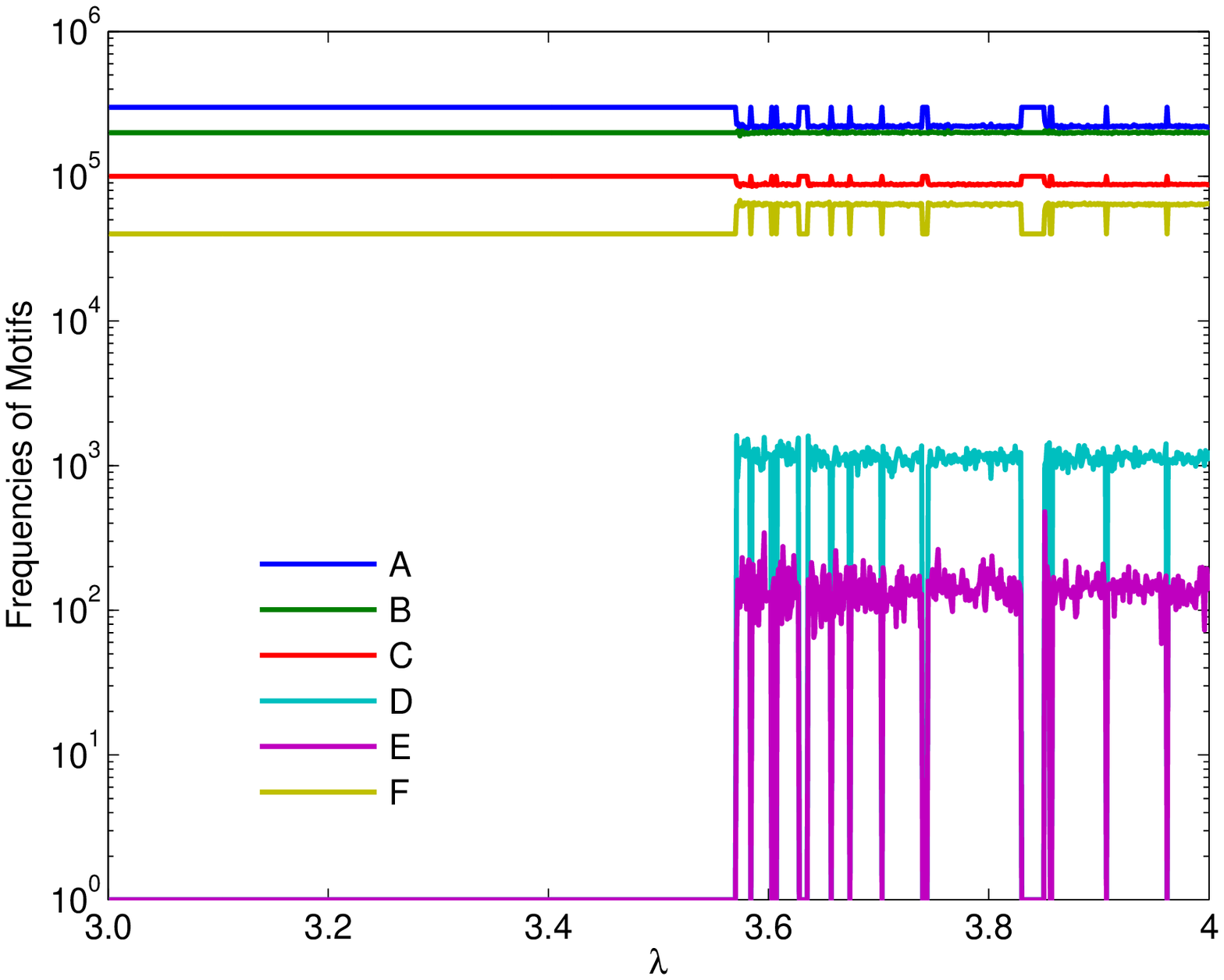}
\put(12,10){(b)}
\end{overpic}
}\\
\parbox{8cm}{\caption{\label{fig:Motiflogsort0}
(a) Motif frequencies from the data sets numerically when tuning the control parameter $\lambda$. Step size is set to be 0.001 here. (b) The same plot in logarithm scale.
}}
\end{figure}

We apply the motif analysis to the full bifurcation picture with $\lambda$ ranging from $3$ to $4$ with a step size equals to $0.001$, as shown in Fig.~\ref{fig:Motiflogsort0}. In all the cases, the dominated motifs are $A$, $B$, $C$ and $F$. Just a few motifs of type $D$ and $E$ as the system enters the chaotic zone, implying the attractors of chaotic data become more heterogeneous. Several periodic windows can be found as the system enters the chaotic zone in which the motif frequencies are very similar to that in regular networks.

\begin{figure}[!ht]
 \centering
\includegraphics[trim=0 0 0 0, clip=true, width=0.4\textwidth]{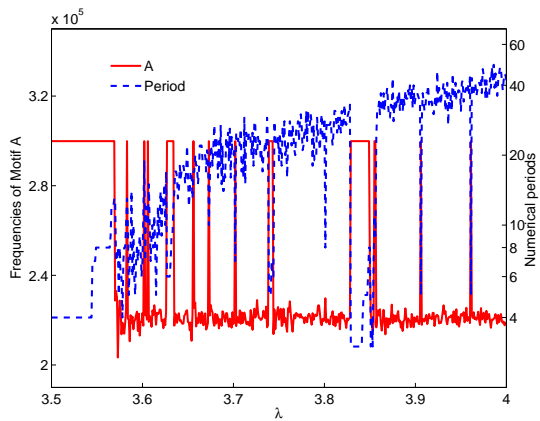}\\
\parbox{8cm}{\caption{\label{fig:Motifsort}
A comparison between frequencies of motif A and numerical period detected from the data sets  when tuning the control parameter $\lambda$. Step size is set to be 0.001 here.}}
\end{figure}

  As a comparison, we plot the frequencies of motif $A$ and the numerical period detected from the data sets in the same graph, as shown in Fig.~\ref{fig:Motifsort}. A large period could be an indicator for chaos, due to the precision in these in numerical calculations. We can conclude from the graph that motifs can be used to describe networks from logistic map of different dynamical regimes.

\subsection{R\"{o}ssler system}
Rather than the map data, we also apply the motif analysis to the R\"{o}ssler system --- a typical continuous dynamical system exhibit period doubling route to choas. The data sets that we investigate are from the $x$ component of the R\"{o}ssler system by solving $x'=-(y+z)$, $y'=x+ay$, $z'=b+(x-c)z$ with $a=0.1$, $b=0.1$ and a tunable $c$, together with a fixed sampling rate of $h=0.1$. The embedded dimension $d_e$ is set as $d_e=10$ that is sufficiently large to study the topological structure and $\tau=10$ which is around the first minimum of the mutual information of the data. We assign each points with $k=4$ nearest neighbors to get the phase space networks with $10000$ nodes, following the modified Xu's methods. Low-period periodic, high-period periodic and chaotic data can be obtained by tuning $c$ accordingly as shown in Table~\ref{table:criticalpoints1}.

\begin{table*}[!h]
  \caption{\label{table:criticalpoints1}
R\"{o}ssler system with different $c$.}
\begin{center}
\begin{tabular}[c]{llc}\hline\hline
    &$c$  &Description \\\hline
P2       &$c=6$                    &Period=2\\
P3       &$c=12$                   &Period=3\\
P4       &$c=8.5$                  &Period=4\\
P6       &$c=12.6$                 &Period=6\\
P8       &$c=8.7$                  &Period=8\\
C1       &$c=9$                    &Chaotic\\
C2       &$c=18$                   &Chaotic\\\hline
\end{tabular}
\end{center}
\end{table*}

\begin{figure}[!ht]
\centering
{\begin{overpic}[trim=0 0 0 0, clip=true, width=0.38\textwidth]{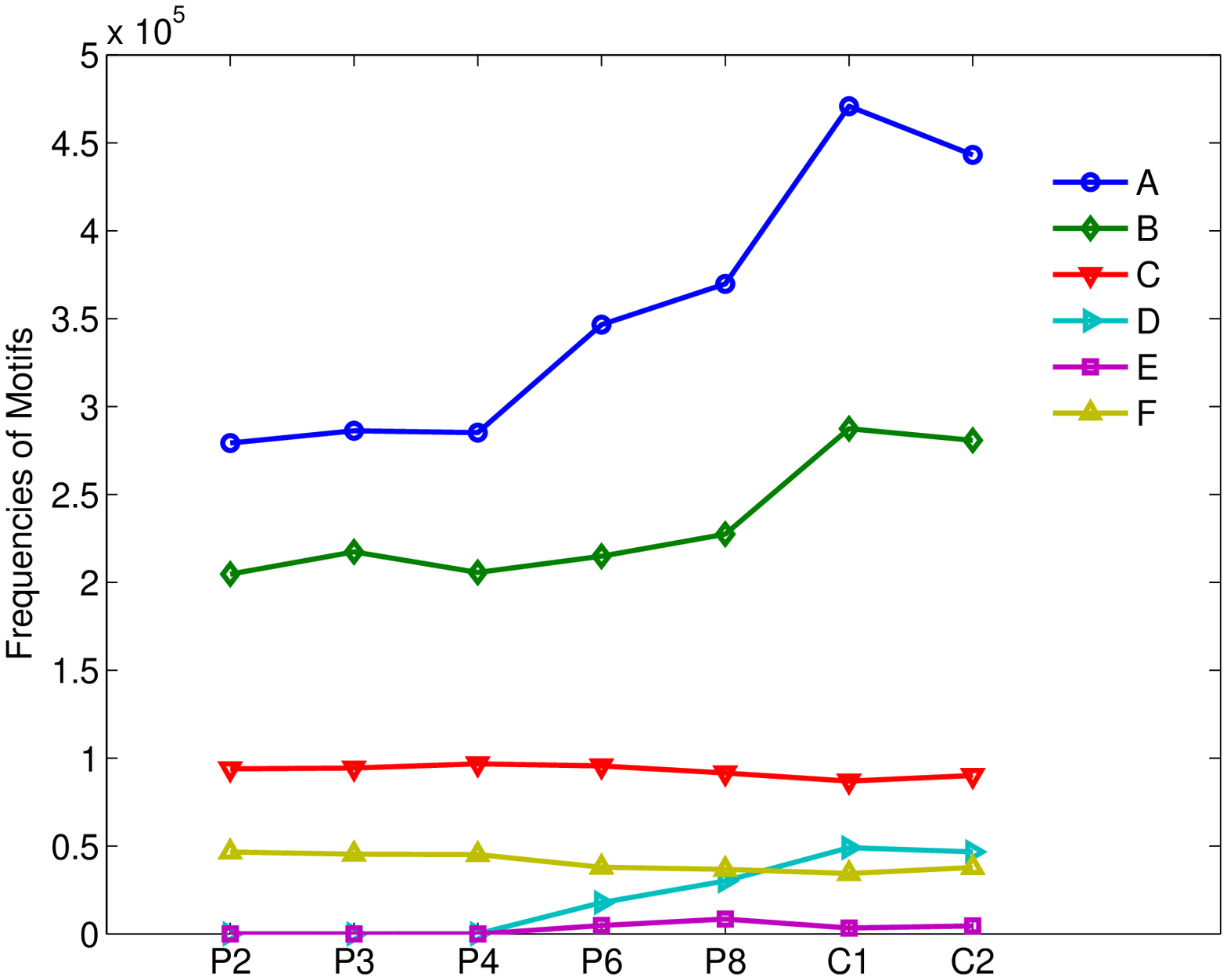}
\put(90,10){(a)}
\end{overpic}\\
\begin{overpic}[trim=0 0 0 0, clip=true, width=0.38\textwidth]{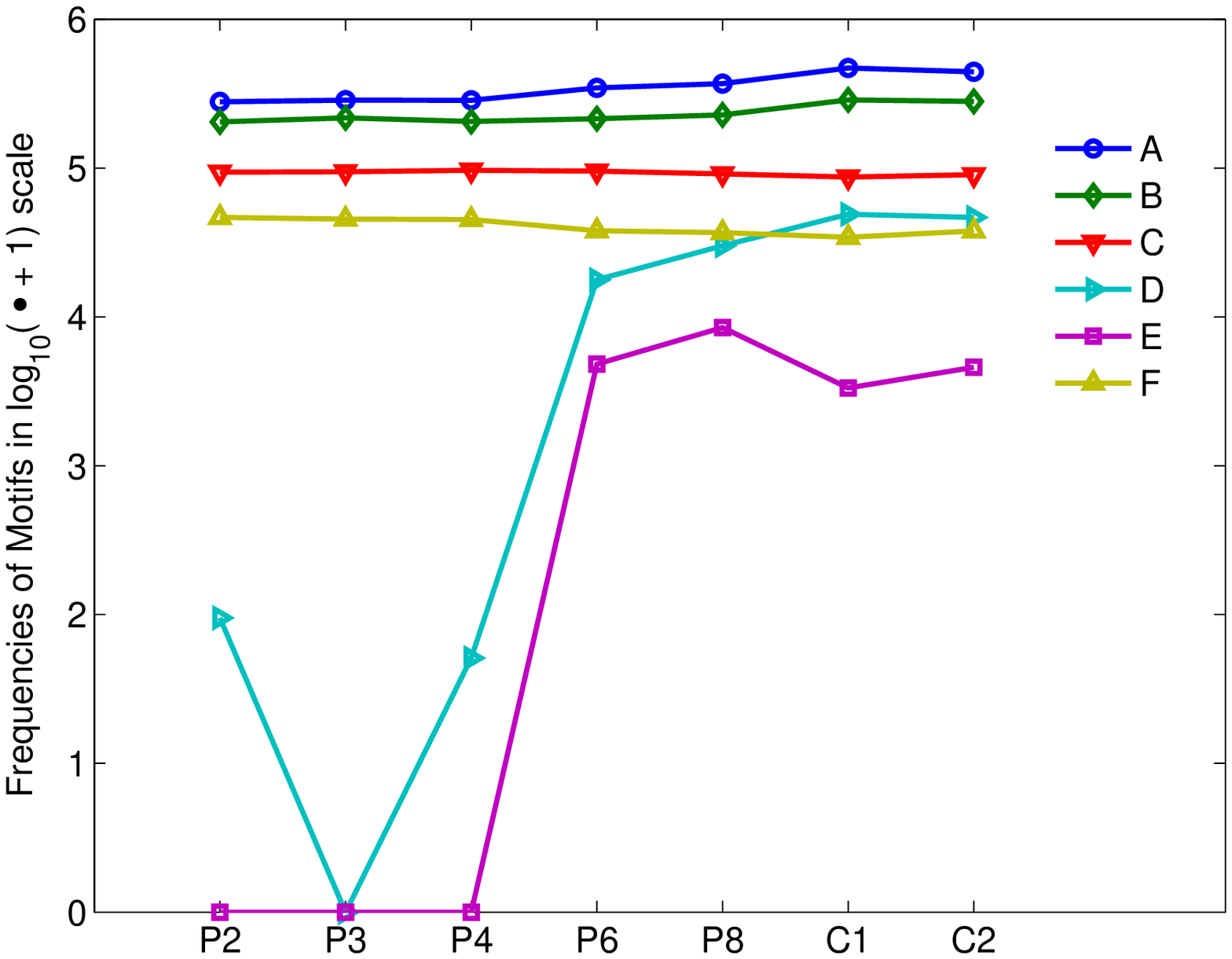}
\put(90,10){(b)}
\end{overpic}
}\\
\parbox{8cm}{\caption{\label{fig:Motifrosbif}
(a) Motif frequencies for periodic and chaotic flow data from R\"{o}ssler system. (b) The same plot in logarithm scale.
 }}
\end{figure}

We measure the motif frequencies for networks from each $c$, as shown in Fig.~\ref{fig:Motifrosbif}. As we expect, Motif $D$ is increased while Motif $F$ is decreased as the system goes from low periodic (P2, P3, P4) to high periodic (P6, P8) and finally to chaos (C1, C2), resulting in the switching of superfamilies from $ABCFDE$ to $ABCDFE$. The absolute motifs frequencies for P2, P3 and P4 are very close to those obtained from periodic logistic map in Sec.~\ref{chap04:logistic}. From periodic to chaotic, we find the increase in motifs of $B$, $D$, $E$ and the decrease in motif $C$, which are in agreement with what happens in the logistic map.  On the contrary, motif $A$ will be more common while motif $F$ will be less common during the transition from low period to chaos, compared with the logistic map. That is because for flow data, usually there is only one network component. Hence, motif frequencies will not be affected by the size of each network component. In this way, motif $F$ will be more common in periodic flow (in which the points are evenly distributed) and correspondingly less common in transitive dynamics (for the distribution of points are non-uniform and thus possesses less mutual coupling).

\begin{figure}[!ht]
 \centering
\includegraphics[trim=0 0 0 0, clip=true, width=0.45\textwidth]{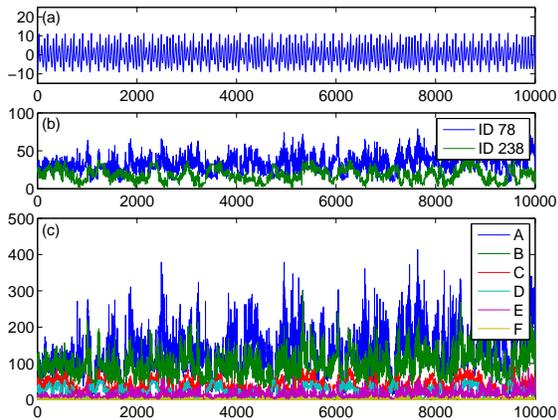}\\
\parbox{8cm}{\caption{\label{fig:rosUPOmotif}
(a) The $x$ component of Chaotic R\"{o}ssler system; (b) Local motif frequencies of order 3; (c) Local motif frequencies of order 4.
}}
\end{figure}

The switching of motif ranks between $D$ and $F$ when the system transits from periodic to chaos relates strongly with the fine scale state recurrence mechanism. A more careful examination of local motifs will provide more details on how the structure of phase space network codes the subtle changing characteristics concerning the chaotic phase space attractor. The data we use is from the chaotic R\"{o}ssler system with $a=0.2$, $b=0.2$, $c=5.7$ and sampling rate $h=0.1$, which starts with an UPO-3 orbit (unstable periodic orbit of order 3) and oscillates with cycle length of around 60 points. The trajectory stays in the vicinity of the UPO-3 for at least 800 points before it is ejected away of the UPO-3. The embedded parameters are selected as $d_e=10$ and $\tau=10$ to build the $k=4$ nearest neighbors phase space network of size $10000$. Figures~\ref{fig:rosUPOmotif}(a) and (b) show the local motif frequencies of order 3 and 4 measured from each data point accordingly. By measuring the local motifs of each network nodes, the data is decomposed into sets of series denoted by motif IDs, 2 for order 3 and 6 for order 4, respectively. We can infer that the distributions of motifs is highly  heterogenous along different orbits, compared with the homogeneity in regular network from a constant data set. The relative proportion of each motif type changes along the trajectories as well.

\begin{figure}[!ht]
 \centering
\includegraphics[trim=0 0 0 0, clip=true, width=0.45\textwidth]{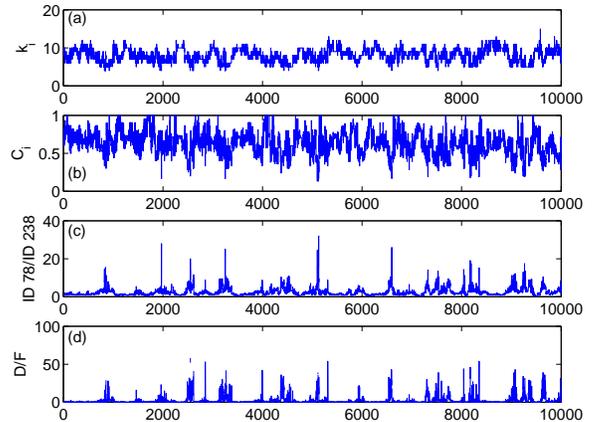}\\
\parbox{8cm}{\caption{\label{fig:rosUPOdcccmotif34}
(a) The local degrees $k_v$; (b) The local clustering coefficients $C_v$; (c)The ratio in frequencies between two types of motifs of order 3, i.e. ID78 and ID238. (d) The ratio in frequencies between motifs $D$ and $F$.
}}
\end{figure}

What we are most interested in are the ratios between the less-connected motif and the most-connected ones which plays the most important role in distinguishing signals of different dynamics into different ``superfamily'' groups. Regarding the motifs of order 3, the less-connected type of motifs is labeled as ID78 while the most connected ones are ID238. For motif of order 4, what we are most interested are the less-connected motifs is motif $D$ rather than the dominated motif $A$, while the most-connected one is motif $F$. In Fig.~\ref{fig:rosUPOdcccmotif34}, we plot the ratios for motif of order 3 and 4, together with the local degrees and local clustering coefficients for comparison. The starting of the trajectories lies along an UPO-3 in which we find low flat values in motif ratio, denoting that the points are densely but homogenously lie along UPO-3. We find peaks as the trajectory is ejected away from UPO-3, leading to the non-uniformnity in points distribution and hence less-connected motifs are more likely to appear while the fully-connected motifs are less likely to emerge.

\section{\label{sec5:con}Conclusions}
In this work we have explored the prevalence of different motif superfamilies as chaotic dynamical systems (both flow and map) undergo a period doubling cascade. We find that with the logistic map, the frequencies of motifs changes accordingly towards different dynamics, mainly because the networks are broken into small components according to different data sets and the size effects in network components affect the motif frequencies of different dynamics. For R\"{o}ssler flow data, usually, there is only one network components, hence rather than the size of network but the network topologies that codes in the existence of the UPOs in the phase space attractor and exhibit different the local motif information along different trajectories. This work provides a significance extension of the techniques introduced by Xu and colleagues~\cite{Xu:Superfamily:2008}. Whereas Xu examined motif superfamily phenomena for motifs of size 4, we have refined the computational techniques that are able to examine motifs of other orders, while local phase space information at a large scale is needed.  Moreover, we have introduced a new network construction refinement that allows for pseudo-periodic and periodic signals to be correctly represented in the network domain. This refinement --- together with the more efficient computational techniques --- allows us to fully explore the evolving period doubling cascade and onset of chaos. Finally, by exploring this well studied dynamical transition we have also been able to better understand the mechanisms underlying the prevalence of different network motifs. Hence, for the first time, we provide a description of {\em why} certain motifs are more representative of different dynamical regimes --- thereby providing a theoretical justification for the empirical observations of Xu {\em et al}~\cite{Xu:Superfamily:2008}.

The novel contribution of this work include: (1) the development of new algorithms for motif characterization of complex networks, (2) a deeper exploration of motif frequency and the connection to state-space (i.e. what features in state-space lead to which motifs), (3) improvements to network construction algorithm that allow for better representation of state space features in the network.

\begin{acknowledgments}
This work was supported by a Hong Kong Polytechnic University direct allocation (G-YG35).
\end{acknowledgments}
\bibliography{ref}
\end{document}